\documentclass{ws-rv961x669}
\usepackage{ws-rv-van}     % numbered citation/references (default)
\usepackage{ws-rv-thm}     % comment this line when `amsthm / theorem / ntheorem` package is used
\usepackage{ws-index}     

\makeindex

\begin{document}

\chapter[Elliptic Flow and the Nuclear Equation of State]{Elliptic Flow and the Nuclear Equation of State\label{ra_ch1}}

%\chapter[Elliptic Flow and the Equation of State]{Elliptic Flow and the Equation of State of Symmetric and Asymmetric Nuclear Matter\label{ra_ch1}}

%\author[Trautmann and Wolter]{Trautmann and Wolter\footnote{Author footnote.}}


%%%%%%%%%%%%%%%%%%%%%%%%%%%%%%%%%%%%%%%%%%%%%%%%%%%%%%%%%%%%%%%%%%%%

%\author{W. TRAUTMANN\footnote{Dedicated to the Memory of Professor Walter Greiner}}

\author{W. TRAUTMANN and H. H. WOLTER}

\address{GSI Helmholtzzentrum f\"{u}r Schwerionenforschung GmbH\\
Planckstr. 1, D-64291 Darmstadt, Germany\\
w.trautmann@gsi.de}

%\author{H. H. WOLTER}

\address{Fakult\"{a}t f\"{u}r Physik, Universit\"{a}t M\"{u}nchen\\
Am Coulombwall 1, D-85748 Garching, Germany\\
hermann.wolter@lmu.de}

%\maketitle

\begin{abstract}
New constraints for the nuclear equation of state at suprasaturation densities have been obtained by measuring collective particle 
flows in heavy-ion reactions at relativistic energies. Ratios and differences of neutron and hydrogen flows in $^{197}$Au + $^{197}$Au 
collisions at 400 MeV/nucleon were used in studies of the asymmetric-matter equation of state. The comparison with predictions of 
transport models favors a moderately soft to linear density dependence, consistent with ab-initio nuclear matter theories. 
Model predictions suggest that comprehensive data sets collected at higher bombarding energies will provide information on 
the asymmetric-matter equation of state in the density range up to two or three times the saturation value. 
\end{abstract}

\body
\section{Introduction}

Collective nuclear motion has always been a topic followed and advanced by Walter Greiner.
His seminal papers on nuclear collective excitation from the 60s of last century 
have guided a generation of nuclear structure physicists.\cite{greiner1,greiner2} The investigation and description
of density oscillations of small amplitude have not only enhanced our knowledge about the collective 
degrees of freedom of atomic nuclei but have also revealed fundamental properties of nuclear matter. 
The nuclear compressibility is of importance for many phenomena in nuclear structure and nuclear
reactions as well as for astrophysics. 

To go beyond the small density interval probed with giant resonances requires nuclear reactions. 
In their famous shock-wave paper of 1974, Scheid, M\"{u}ller and Greiner have studied new phenomena 
produced by the collective pressure in the shock-compressed overlap zone.\cite{greiner4} They have shown that heavy-ion reactions 
at sufficiently high energy provide us with the possibility to compress nuclear matter up to several times the saturation 
density $\rho_0$ encountered in the interior of heavy nuclei. The properties of nuclear matter 
at suprasaturation densities may thus be studied in laboratory experiments.\cite{dani02} From the extensive search for observables 
suitable for probing the brief high-density phase of the collision, collective flows and sub-threshold 
production of strange mesons have appeared as most useful. A consensus has been reached that a soft equation of state (EoS), 
corresponding to a compressibility $K \approx 200$~MeV and including momentum dependent interactions, best describes the
high-density behavior of symmetric nuclear matter. It is based on studies of flow and kaon 
production within the framework of transport theory.\cite{dani02,sturm01,fuchs01}. This finding has very recently been confirmed in a 
new analysis of the high-precision flow data measured by the FOPI Collaboration at the GSI laboratory.\cite{reisdorf12,lefevre16} 
We may note here that the conclusion that $K$ should be of the order of 200 MeV was presented by Greiner and coworkers 49 years ago, 
based on a theoretical analysis of the elastic-scattering excitation function for $^{16}$O + $^{16}$O measured by the Yale group.\cite{greiner3}

In recent years, motivated by the impressive progress made in observing properties of neutron 
stars and in understanding details of supernova explosion scenarios, the EoS of neutron-rich
asymmetric matter has received increasing attention.\cite{lattprak07,lipr08,ditoro10,EPJA2014} 
The symmetry energy, equal to the difference between the energies per nucleon of neutron matter and of symmetric 
matter, is seen as one of the biggest unknowns in this context. We have precise information 
on the symmetry energy near saturation density from the knowledge of nuclear masses.
For densities below saturation, investigations of nuclear structure phenomena and of heavy ion
reactions in the Fermi energy regime have constrained the symmetry energy
considerably.\cite{tsang09,tsang12,horo14}. The importance of clustering at subsaturation
densities has recently been demonstrated\cite{horo06} and confirmed in comparison with 
experimental data from intermediate-energy reactions.\cite{nato10,typel10,qin12}
At suprasaturation density, however, the symmetry energy is still largely unknown for several 
reasons. Phenomenological forces, even though well constrained near saturation, yield 
largely diverging results if they are extrapolated to higher densities.\cite{brown00,fuchs06}
Many-body calculations with realistic potentials face the problem that three-body forces and 
short-range correlations are not sufficiently constrained at the higher densities 
at which their importance increases.\cite{subedi08,xuli10,steiner12,hen14sc} 
Even the magnitude of the kinetic contribution is possibly
modified by a redistribution of nucleon momenta due to short-range correlations in 
high-density nuclear matter.\cite{carb11,hen15,yong16}

\begin{figure}[!htb]   %Fig. 1
\centerline{\includegraphics[angle=270,width=0.52\columnwidth]{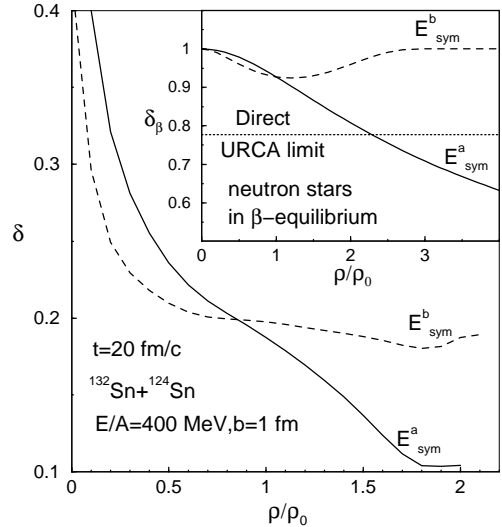}}
\vskip -0.1cm  
  \caption{Isospin asymmetry $\delta = (\rho_n -\rho_p)/\rho$ as a function of the 
normalized density $\rho/\rho_0 $ %over the whole space 
at time $t = 20$~fm/$c$ in the $^{132}$Sn + $^{124}$Sn reaction with 
a stiff ($E^a_{\rm sym}$) and with a soft ($E^b_{\rm sym}$) density dependence of the 
nuclear symmetry energy. 
The corresponding correlation for neutron stars in $\beta$-equilibrium is shown in
the inset (reprinted with permission from Ref.~36; 
Copyright (2002) by the American Physical Society).}

\label{fig:li_prl02fig2}
\end{figure}

The symmetry energy appears in nearly every aspect of nuclear structure and reactions and, 
as a consequence, a variety of constraints obtained with different methods have become 
available.\cite{lihan13} However, quantities as, e.g.,
the thickness of the neutron skin in heavy nuclei or the isospin transport 
in reactions of isospin-asymmetric nuclei in the Fermi-energy domain are predominantly 
sensitive to the strength of the symmetry term at densities near or below 
the saturation value.\cite{lipr08,tsang09,tsang12,baran05,klimk07,tamii11,zhangchen13}
The example shown by Brown\cite{brown13} demonstrates that a very precise knowledge of 
the neutron-skin thickness of $^{208}$Pb will be required if an extrapolation from density $\rho \approx 0.6\rho_0$,
where the symmetry energy is well determined,
to the saturation density $\rho_0$ is attempted. This emphasizes the need for more direct high-density probes.
As in the case of symmetric nuclear matter, collective flows and sub-threshold particle
production are obvious candidates.

A strong motivation for exploring the information contained in isotopic flows was provided some time ago
by Bao-An Li when he pointed to the parallels in the density-dependent isotopic compositions 
of neutron stars and of the transient systems formed in collisions of neutron-rich nuclei 
depending on the EoS input used in the calculations.\cite{li02} 
Figure~\ref{fig:li_prl02fig2} illustrates the remarkable fact that 
the same physical laws govern objects differing by 18 orders of magnitude 
in linear scale or 55 orders of magnitude in mass. Properties of 
exotic astrophysical objects may thus be inferred from data obtained in laboratory experiments,
and vice versa. 
A main difficulty resides in the comparatively small asymmetry of atomic nuclei available for experiments.
Symmetry effects are, therefore, always small relative to those of the dominating isoscalar forces.
A partial cancellation of the latter may be expected in differences or ratios of observables between isotopic partners.

A further encouragement was provided by transport model calculations indicating that
the elliptic flows of free neutrons and free protons respond differently
to variations of the parametrization of the symmetry energy.\cite{russotto11}
Elliptic flow refers to the second Fourier component of the azimuthal anisotropy of particle emissions.
It has motivated a reanalysis of the FOPI-LAND data for $^{197}$Au + $^{197}$Au collisions 
at 400 MeV/nucleon, collected many years ago and used to demonstrate the existence of 
neutron squeeze-out in this energy regime.\cite{leif93,lamb94} 
Squeeze-out refers to a dominant out-of-plane emission of particles, relative to in-plane emission, 
and is considered as evidence for the pressure buildup in the collision zone. 
It should therefore be particularly sensitive to the high-density EoS.
The analysis has favored a moderately soft to linear 
density dependence of the symmetry energy.\cite{russotto11,rev_epja}

This finding had a particular significance, in spite of a large statistical uncertainty. 
Rather different conclusions, ranging from a super-soft to a super-stiff behavior of the symmetry energy, 
had previously been reached in analyses of the $\pi^-/\pi^+$ production ratios, measured by the
FOPI Collaboration\cite{reis07}  for the same $^{197}$Au + $^{197}$Au reaction,
with different transport models.\cite{ferini05,xiao09,feng10,xie13} 
In particular, the super-soft result, first presented by Xiao {\it et al.},\cite{xiao09} has initiated 
a broad discussion of how it might be reconciled with other observations as, e.g., 
properties of neutron stars.\cite{xiao09,baoan11,wen09} 
The FOPI-LAND elliptic-flow data were found to be inconsistent with this extreme 
assumption and, in fact, fairly independent of particular choices made for the model parameters 
used for the quantum-molecular-dynamics (QMD) transport-model calculations.\cite{russotto11,cozma11} 

The obvious need to improve the statistical accuracy beyond that of the existing
data set has initiated a dedicated measurement by the ASY-EOS Collaboration of
collective flows in collisions of $^{197}$Au + $^{197}$Au as well as of $^{96}$Zr + $^{96}$Zr 
and $^{96}$Ru + $^{96}$Ru. It was carried out in 2011 at the GSI laboratory with the LAND\cite{LAND} detector 
coupled to a subset of the CHIMERA\cite{CHIMERA} detector array.\cite{russotto16}
In the following sections, the present situation will be described in relation to the new 
results from the ASY-EOS experiment.

\section{Present Knowledge of the Symmetry Energy}

Our knowledge of the symmetry energy is originally based on nuclear masses whose dependence 
on the isotopic composition is reflected by the symmetry term in the Bethe-Weizs\"{a}cker 
mass formula. A density dependence is already indicated by the use of separate bulk and 
surface terms in refined mass formulae. In the Fermi-gas model, it is given by a proportionality
to $(\rho / \rho_0)^{\gamma}$ with an exponent $\gamma = 2/3$,
where $\rho_0 \approx 0.16~{\rm nucleons/fm}^{3}$ is the saturation density.
This kinetic contribution to the symmetry energy, however, amounts to only about 1/3 of the symmetry term
of approximately 30 MeV for nuclear matter at saturation. The major contribution
is given by the potential term reflecting properties of the nuclear forces.

\begin{figure}[!htb]   %Fig. 2
\centerline{\includegraphics[width=0.75\columnwidth]{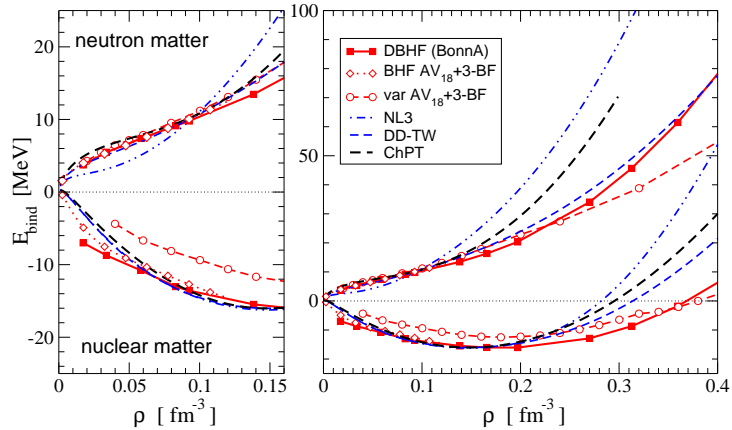}}
\vskip -0.1cm  
  \caption{EoS in nuclear matter and neutron matter as a function of density.  
BHF/DBHF and variational calculations with realistic forces are compared to  
phenomenological density functionals NL3 and DD-TW and to 
ChPT. The left panel zooms the low density range  
(from Ref.~22, 
reprinted with kind permission from Springer Science+Business Media).
}
\label{fig:fuchs06}
\end{figure}

\begin{figure}[!htb]   %Fig. 3
\centerline{\includegraphics[width=0.75\columnwidth]{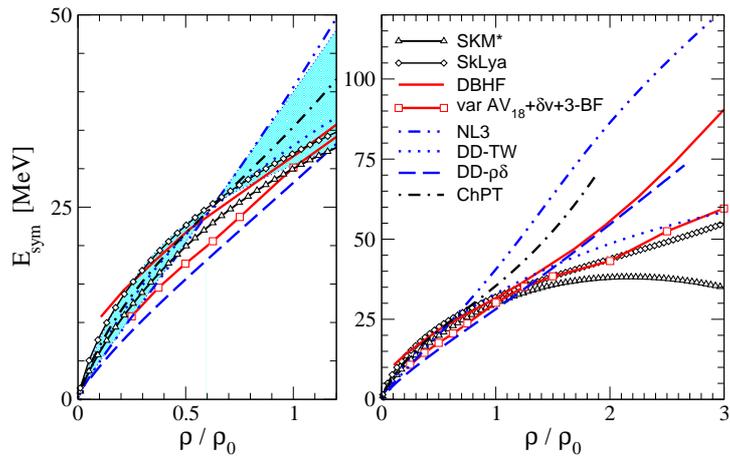}}
\vskip -0.1cm  
  \caption{Symmetry energy as a function of density as predicted by different  
models. The left panel zooms the low density range up to saturation. The full lines
represent the DBHF and variational approaches using realistic forces   
(from Ref.~22, 
reprinted with kind permission from Springer Science+Business Media).
}
\label{fig:fuchs06b}
\end{figure}

Nuclear many-body theory has presented us with a variety of predictions for the nuclear
equation of state.\cite{dani02,lipr08,fuchs06,baldo04,fukukawa15,drischler16} 
The examples shown in Fig.~\ref{fig:fuchs06} for the two cases of 
symmetric nuclear matter and of pure neutron matter demonstrate that, overall, the results
are quite compatible among each other, except for densities exceeding saturation at 
which the predictions diverge. 
The symmetry energy $E_{\rm sym}$ can be defined as the coefficient of the quadratic 
term in an expansion of the energy per particle in the asymmetry  
$\delta = (\rho_n-\rho_p)/\rho$, where $\rho_n, \rho_p,$ and $\rho$
represent the neutron, proton, and total densities, respectively,
\begin{equation}
E/A(\rho,\delta) = E/A(\rho,\delta = 0) + E_{\rm sym}(\rho)\cdot \delta^2 + \mathcal{O}(\delta^4).
\label{eq:e_sym}
\end{equation}

\noindent In the quadratic approximation, the symmetry energy is the difference 
between the energies of symmetric matter ($\delta = 0$) and neutron matter ($\delta = 1$). It is shown in Fig.~\ref{fig:fuchs06b} for a similar range of models as in Fig.~\ref{fig:fuchs06}. 
Also the symmetry energy diverges at high density, as expected from 
Fig.~\ref{fig:fuchs06}, while most empirical models coincide near or slightly below saturation, 
the density range at which constraints from finite nuclei are valid.

In calculations using realistic forces fitted to two- and three-nucleon data, the 
uncertainty is mainly related to the short-range behavior of the nucleon-nucleon force
and, in particular, to the three-body and tensor forces.\cite{subedi08,xuli10,steiner12,hen14sc} 
The three-body force has been shown to make an essential but quantitatively small contribution 
to the masses of light nuclei.\cite{wiringa02} 
The extrapolation of the partly phenomenological terms used there to higher densities is, 
however, highly uncertain.\cite{xuli10} 
The general effect of including three-body forces in the calculations is a 
stiffening of the symmetry energy with increasing density.\cite{burgio08,hebe10}
Short-range correlations become also increasingly important at higher densities.
Results from very recent new experiments will, therefore, have a 
strong impact on predictions for high-density nuclear matter.\cite{subedi08,hen14sc,carb11,hen15,yong16}
Data from neutron-star observations provide important constraints already now.\cite{steiner10} 
With the improvement of observational methods, they will become more stringent in the near future.\cite{NICER} 

At higher energies, the momentum dependence of the nuclear forces
becomes important.\cite{lipr08,ditoro10,giordano10,feng12} 
It is well known that nuclear mean fields are momentum dependent, as seen, e.g., 
in the energy dependence of the nuclear optical potential. The dominating effect 
is in the isoscalar sector but there is also an important isovector momentum
dependence. It manifests itself as an energy dependence of the isospin-dependent part
of the optical potential
but can also be expressed in terms of a difference of the effective masses of protons and
neutrons.\cite{fuchs06}. Even the ordering of these effective masses is still an open
problem.\cite{lihan13,liguoli15,coupland16} It has, moreover, been shown that the 
effective mass differences and the asymmetry dependence of the EoS are both influencing
particle yields and flow observables, and that additional observables will be needed
to resolve the resulting ambiguity.\cite{ditoro10,giordano10,feng12,zhang14}

\begin{figure}[!htb]   %Fig. 4
\centerline{\includegraphics[width=0.72\columnwidth]{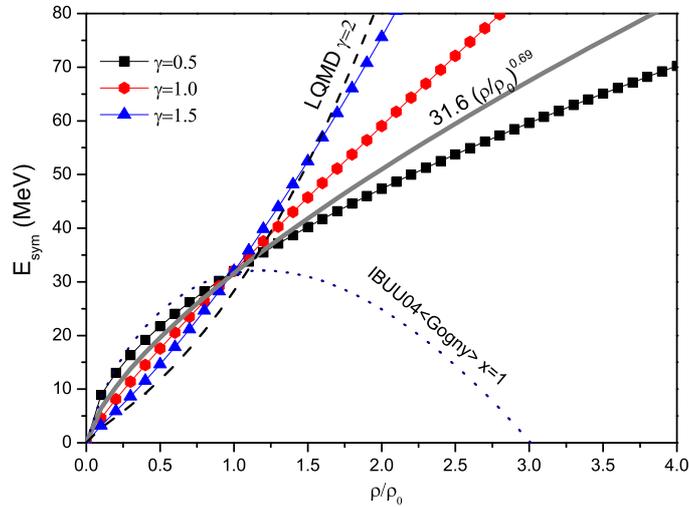}}
\vskip -0.1cm  
  \caption{Parametrizations of the nuclear symmetry energy as used in transport codes: 
three parametrizations of the potential term used in the 
UrQMD (Ref.~65) 
with power law coefficients $\gamma = 0.5, 1.0$, and 1.5 (lines with symbols as indicated), 
the result with $\gamma = 0.69$ obtained from analyzing
isospin diffusion data with the IBUU04 (full line, Ref.~67),
and the super-soft and stiff parametrizations obtained from analyzing the $\pi^-/\pi^+$ 
production ratios with the IBUU04 (dotted line, Ref.~43) and the LQMD 
(dashed line, Ref.~44) transport models
(from Ref.~69,
reprinted with kind permission from Springer Science+Business Media).
}
\label{fig:params}
\end{figure}

Transport theories needed for calculating the temporal evolution of nuclear reactions
often use simplified descriptions of the composition-dependent part of 
the nuclear mean field.  
In the UrQMD of the group of Li and Bleicher,\cite{qli06} 
the potential part of the symmetry energy is defined with two parameters, 
the value at saturation density, usually taken as 22 MeV in their calculations, 
and the power-law coefficient $\gamma$ describing the dependence on
density, %as $(\rho / \rho_0)^{\gamma}$,
\begin{equation}
E_{\rm sym} = E_{\rm sym}^{\rm pot} + E_{\rm sym}^{\rm kin} 
= 22~{\rm MeV} \cdot (\rho /\rho_0)^{\gamma} + 12~{\rm MeV} \cdot (\rho /\rho_0)^{2/3}.
\label{eq:pot_term}
\end{equation}

\noindent In other codes the nuclear potential of Das {\it et al.} 
with explicit momentum dependence in the isovector sector is used.\cite{das03} 
There, as in the Boltzmann-Uehling-Uhlenbeck-type code IBUU04 developed by the groups of Li and Chen,\cite{lipr08,lichen05} 
the density dependence of the symmetry energy is characterized by a parameter $x$ 
appearing in the potential expressions.
Examples of these parametrizations and of results obtained from the analysis of experimental
reaction data are given in Fig.~\ref{fig:params}. The stiff ($E^a_{\rm sym}$) and 
soft ($E^b_{\rm sym}$) density dependences of Fig.~\ref{fig:li_prl02fig2} correspond 
approximately to the cases $\gamma = 1$ and $x = 1$ shown here. Density functionals of the Skyrme type are also increasingly used 
in transport model calculations.\cite{coupland16,wang14}

Parametrizations of this kind have the consequence that, once the symmetry energy at the 
saturation point is fixed, a single value at a different density or, alternatively, 
the slope or curvature at any density will completely determine the 
parametrization. Measurements of a variety of observables in nuclear structure and reactions 
have, in this manner, been used to obtain results for the density dependence of the 
symmetry energy. They are conventionally
expressed in the form of the parameter $L$ which is proportional to the slope 
at saturation, 
\begin{equation}
L = 3\rho_0 \cdot dE_{\rm sym}/d\rho |\rho_0.
\label{eq:l}
\end{equation} 

\noindent Most results with their errors fall into the interval 20~MeV $\le L \le$ 100~MeV 
and are compatible with a most probable value $L\approx 60$~MeV, roughly 
corresponding to a power-law coefficient $\gamma = 0.6$.\cite{lipr08,tsang09,tsang12,lihan13,baoan11,lattlim13} 
The full line in Fig.~\ref{fig:params} represents, e.g., the result deduced by Li and Chen
from the MSU isospin-diffusion data and the neutron-skin thickness in $^{208}$Pb.\cite{lichen05}
The corresponding slope parameter is $L=65$~MeV. 
Rather similar constraints have been deduced from recent investigations and observations 
of neutron-star properties.\cite{steiner12,hebe10,steiner10}

Considerable progress regarding the correlation of the symmetry energy
with particular observables in different models
has been made by the Florida and Barcelona 
groups.\cite{todd_piek05,rocamaza11} In continuation of the work of Typel and Brown\cite{typel01}
and of Furnstahl,\cite{furn02} a universal 
correlation between the thickness of the neutron-skin of $^{208}$Pb and the $L$ parameter 
has been found for empirical mean-field interactions.\cite{rocamaza11} 
The determination of the neutron-skin thickness by measuring the 
parity-violating contribution to electron scattering at high energy will thus offer a practically
model-free access to the slope at saturation.\cite{horo12,PREX12}

\section{Probing High Densities}

Densities of two to three times the saturation density may be reached on time scales 
of $\approx$ 10--20~fm/$c$ in the central zone of heavy-ion collisions at
relativistic energies of up to $\approx 1$~GeV/nucleon.\cite{li_npa02,junxu13} 
The resulting pressure produces a collective outward motion of the compressed 
material whose strength will be influenced by the symmetry energy in asymmetric 
systems.\cite{dani02,greco03} At the same time, the excitation of $\Delta$ resonances in hard 
nucleon-nucleon scatterings leads to the production and subsequent emission of 
charged and neutral $\pi$ and $K$ mesons. 
Their property as messengers from the early reaction phase identifies them as potential 
probes for the high-density symmetry energy. Calculations show that the highest 
sensitivities may be expected at near-threshold energies.\cite{fuchs01,ferini05,xlopez07} 
In both cases, collective flows or meson production, isotopic differences or ratios of 
observables are useful to minimize the sensitivity to the isoscalar part in the EoS while 
maximizing that to the symmetry term.

The FOPI Collaboration has collected an extensive set of $\pi^-/\pi^+$ 
production ratios measured for the four reactions $^{40}$Ca + $^{40}$Ca, $^{96}$Zr + $^{96}$Zr, $^{96}$Ru + $^{96}$Ru, and 
$^{197}$Au + $^{197}$Au at several energies between 0.4 and 1.5 GeV/nucleon.\cite{reis07} 
Theoretical analyses of these data, however, have come to
conflicting conclusions, suggesting everything from a rather stiff to a
super-soft behavior of the symmetry energy.\cite{ferini05,xiao09,feng10,xie13} 
The reason may lie in the treatment of the pion in-medium and $\Delta$ dynamics\cite{junxu13,cozma16,ikeno16} and in competing effects 
of the mean fields and the $\Delta$ thresholds\cite{ferini05,song15} which will
require further studies.\cite{baoan15} 
Cozma has demonstrated very recently that both types of observables, 
elliptic flow and pion production, can lead to compatible constraints for the stiffness 
of the symmetry energy, provided that care is taken to conserve the energy in inelastic collisions producing a $\Delta$ resonance.\cite{cozma16} 
It may be done at a local or global level by accounting for the potential energy of hadrons in the treatment of two-body collisions.
The sensitivity inherent in the spectral dependence of pion production has been pointed out by several 
authors.\cite{hong14,cozma17,tsang17} 
New information is expected to come from measurements performed by the 
S$\pi$RIT Collaboration\cite{tsang17,sprit} at RIKEN with radioactive Sn beams.

\section{Transport Theory}

Common to all high-density probes is the need to relate the observations made for the asymptotic outcome of the reaction
to the short interval during which the compressed matter exists and impacts the subsequent evolution. 
Heavy-ion collisions are non-equililibrium processes and non-equilibrium theory is, therefore, needed for their interpretation. Hydrodynamical or statistical descriptions may be useful for describing certain stages of the reaction process. A unified description requires transport theory to follow the evolution from the initial state of the collision system to the final stage when the strong interactions cease to act between its components. Such approaches have first been developed in the group of Walter Greiner and at Michigan State University in the 80s,
\cite{bertsch84,kruse85,aichelin86} 
and have since evolved into a very valuable tool for extracting physics information from heavy-ion experiments. 

Two families of transport approaches with different philosophies have been developed in nuclear physics. Those of the Boltzmann-Vlasov type (often named Boltzmann-Uehling-Uhlenbeck (BUU) approaches) decribe the evolution in time of the one-body phase space density $f(\vec{r},\vec{p};t)$ under the influence of a mean field $U[f]$ and of two-body collisions
\begin{eqnarray}
& &\frac{\partial f_1}{\partial t} +  \frac{\vec{p}}{m} \nabla_{\vec{r}} f_1 - \nabla_{\vec{r}} U\nabla_{\vec{p}} f_1 =\\ 
& & (\frac{2\pi}{m})^3 \int{ d\vec{p}_2 d\vec{p}_3 d\vec{p}_4
|\vec{v}_1-\vec{v}_2| \sigma_{NN}(\Omega_{12})  \delta(\vec{p}_1+\vec{p}_2-\vec{p}_3-\vec{p}_3)
(f_3 f_4 \bar{f}_1 \bar{f}_2 -f_1 f_2 \bar{f}_3 \bar{f}_4) }. \nonumber
\end{eqnarray}
\label{eq:kineq}

\noindent Here $f_i= f(\vec{r},\vec{p_i};t), \bar{f}_i=(1-f_i),$  $v_i$ are velocities, and $\sigma_{NN}(\Omega)$ is the in-medium nucleon-nucleon (NN) cross-section. 
The potential $U[f]$ and the cross-section are either derived from
an energy density functional, or are parametrized in order to test them relative to the data.
If particle production, e.g. of pions and $\Delta$ resonances, is to be considered, additional physics input is needed: inelastic cross sections, potentials of the new particles, their cross sections for collisions with other particles and, possibly, the finite mass distributions of instable particles. 

The solution of this equation is achieved with simulations using the test-particle (TP) method where the distribution function is represented in terms of finite elements as
\begin{equation}
f(\vec{r},\vec{p};t)=\frac{1}{N_{TP}} \sum_{i=1}^{AN_\text{TP}} g(\vec{r}-\vec{r}_i(t)) \, \tilde{g}(\vec{p}-\vec{p}_i(t)) ;
\end{equation}
\label{eq:distr}

\noindent here $N_\text{TP}$ is the number of TPs per nucleon, $\vec{r}_i$ and $\vec{p}_i$  are the time-dependent coordinates and momenta of the TPs, and $g$ and $\tilde{g}$ are the shape functions in coordinate and momentum space (e.g.\ $\delta$-functions or Gaussians), respectively.  Upon inserting this ansatz into the l.h.s.\ of Eq.~\ref{eq:kineq}, Hamiltonian equations of motion for the test particles are obtained,
$\frac{\text{d}\vec{r}_i}{\text{d}t}=\vec{\dot p}$ and $\frac{\text{d}\vec{p}_i}{\text{d}t}=-\vec{\nabla}_{r_i} U$. The collision term is simulated stochastically, by performing TP collisions with a probability depending on the cross section and obeying the Pauli principle for the final state according to blocking factors $\bar{f}_i=(1-f_i).$  

In the second family, the quantum molecular dynamics (QMD) model, the evolution of the collision is formulated in terms of the evolution of the coordinates $R_i(t)$ and momenta $P_i (t)$ of individual nucleons, similarly as in classical molecular dynamics, but with particles of finite width representing minimum nucleon wave packets. They move under the influence of NN forces. The method can also be viewed as being derived from the Time-Dependent Hartree (TDH)
method with a product trial wave-function of single-particle states in Gaussian form.
One obtains equations of motion of the same form as in BUU for the coordinates of the wave packets. 
Also a stochastic two-body collision term is introduced and treated in very much the same way as in BUU, but now  
for nucleons and the full NN cross section. There are also relativistic formulations for both approaches using 
relativistic density functionals. Of the codes used in the analysis of the flow data discussed here UrQMD and TuQMD 
are relativistic codes, IQMD is non-relativistic. A review of the BUU method is given in Ref.~92 
while the QMD method is reviewed in Ref.~93. 

The main difference between the two approaches lies in the amount of fluctuations and correlations in the 
representation of the phase space distribution. In the BUU approach, the phase-space distribution function 
is seen as a smooth function of coordinates and momenta and can be increasingly better approximated by increasing 
the number of TPs. In the limit of $N_\text{TP}\rightarrow \infty$, the BUU equation is solved exactly but the solution 
is deterministic and contains neither fluctuations nor correlations. If they are considered to be important, as is 
the case when looking at cluster and fragment production, they have to be introduced explicitely through the Boltzmann-Langevin 
equation with a fluctuation term on the r.h.s. of Eq.~\ref{eq:kineq}. In the QMD, fluctuations are present 
because of the finite number of wave packets in the representation. In addition, classical correlations are present 
if explicit two-body interactions are used. The fluctuations in QMD-type codes are regulated and smoothed by choosing 
appropriate width parameters of the wave packets. QMD can be seen as an event generator solving the time evolution of 
different events independently. Event-by-event fluctuations are not even suppressed in the limit of infinitely many events. 

The results of simulations with the two methods are thus expected to be similar, though not necessarily identical, as far 
as one-body observables are concerned as, e.g., the flow observables discussed in this article. Larger differences are 
expected for observables depending on fluctuations and correlations, such as the production of clusters and intermediate-mass 
fragments. Generally, the description of observables going beyond the mean field level is a question under active discussion 
in transport theory. In the experiment, copious numbers of light clusters and fragments are observed in heavy-ion collisions,
particularly at lower energies.
%The experimentally observed heavy-ion collisions, in their final states, contain copious numbers of light clusters and, particularly at lower energies, also intermediate mass fragments. 

In addition to these more fundamental differences between existing transport approaches, there are also differences that are 
caused by different implementations of the highly complex transport theories. Analyses of experimental data with 
seemingly similar physics input have lead to rather different conclusions. The analyses of the FOPI $\pi^-/\pi^+$ ratios  
represent an example. In order to reach a better understanding of possible reasons, a code-comparison project has been started. 
In a first publication, results for a standardized heavy-ion collision with identical physics input were compared.\cite{junxu16} 
Eighteen commonly used transport codes, nine of BUU and nine of QMD type, were included. Quantitatively, 
the differences were found to depend on the incident energy and amounted to approximately
30\% at 100 and 15\% at 400 MeV/nucleon, respectively.
The comparison is presently continued with calculations for infinite nuclear matter. 
There the different ingredients of the transport codes can be tested separately and compared to exact limits.

\section{Elliptic Flow}

A "flow of nuclear matter out of the regions of compressed densities" was already predicted by Scheid, Ligensa and 
Greiner 49 years ago when they investigated central collisions of $^{16}$O nuclei.\cite{greiner3} 
The transverse emission of nuclear matter from the compressed interaction zone, the squeeze-out as it has  been termed, has
first been observed in experiments at the Bevalac.\cite{gutbrod90} In a
sphericity analysis, the event shape in three dimensions was characterized by a
kinetic-energy flow tensor whose main orientation with respect to the beam direction
represents the collective sidewards or directed flow.
A difference in the two minor axes indicates the existence of elliptic flow. At the bombarding energies of up to several
GeV/nucleon investigated in these studies, a preferential emission of charged particles
perpendicular to the reaction plane has been observed. 
The shadowing by the spectator remnants reduces the in-plane flow, so that the strength of the
off-plane emission, as quantified by the azimuthal anisotropy, reflects the internal pressure.

It has become customary to express both, directed and elliptic flows, 
and also possible higher flow components by means of a Fourier decomposition
of the azimuthal distributions measured with respect to the orientation of the 
reaction plane $\phi_R$,\cite{ollitrault97,poskanzer98}   
\begin{equation} 
\frac{dN}{d(\phi-\phi_R)} = \frac{N_0}{2\pi} 
\left( 1+2 \sum_{n\geq1} v_n \cos n(\phi-\phi_R)\right),
\label{eq:defvn} 
\end{equation}

\noindent where $N_0$ is the azimuthally integrated yield. The
coefficients $v_{n} \equiv \langle\cos n(\phi-\phi_R)\rangle$ are functions of 
particle type, impact parameter, rapidity $y$, and the transverse momentum $p_t$; 
$v_{1}$ and $v_{2}$ represent the directed and elliptic flows, respectively.

Elliptic flow has become an important observable at other energy regimes as well. At
ultrarelativistic energies, the observation of the constituent-quark scaling of elliptic flow
is one of the prime arguments for deconfinement during the early collision phase,
and the behavior as an almost perfect liquid of the formed strongly interacting quark-gluon plasma are deduced from the observed
magnitude of collective motions.\cite{abelev07,fries08,adare12,heinz13}
It implies that elliptic flow develops very early in the collision which is valid also
in the present range of relativistic energies, as confirmed by calculations.\cite{dani00}
Isotopic flow differences appear thus very suitable for studying
mean-field effects at high density.

\begin{figure}[!htb]   %Fig. 5
\centerline{\includegraphics[width=0.64\columnwidth]{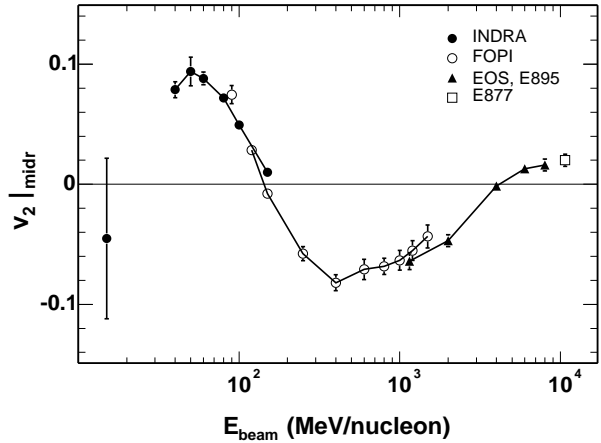}}
\vskip -0.1cm  
\caption{Elliptic flow parameter $v_{2}$ at mid-rapidity for $^{197}$Au+$^{197}$Au 
collisions at intermediate impact parameters ($\approx 5.5-7.5$~fm) as a function of incident
energy. The filled and open circles represent the INDRA
and FOPI data\protect\cite{lukasik05,andronic05} for $Z=1$ particles,
the triangles represent the EOS and E895 data\protect\cite{pinkenburg99} for
protons, and the square represents the E877 data\protect\cite{bmunzinger98}
for all charged particles
(from Ref.~103,
reprinted with kind permission from Springer Science+Business Media).
}
\label{fig:v2corr}
\end{figure}

An excitation function of the elliptic flow of $Z=1$ particles in 
$^{197}$Au+$^{197}$Au collisions, compiled from various experiments,\cite{andro06} 
is shown in Fig.~\ref{fig:v2corr}. Squeeze-out perpendicular to the reaction plane, 
i.e. $v_2 < 0$, as a result of shadowing by the spectator remnants is observed
at incident energies between about 150~MeV/nucleon and 4~GeV/nucleon with a maximum near
400 MeV/nucleon.  At lower energies, the collective angular momentum in the mean-field 
dominated dynamics causes the observed in-plane enhancement of emitted reaction products.
The figure also shows that elliptic flow can be measured quite precisely, as demonstrated by the good agreement of data sets from
different experiments in the overlap regions of the studied intervals in collision 
energy.\cite{reisdorf12,andro06,lukasik05} 

\begin{figure}[!htb]   %Fig. 6
\centerline{\includegraphics[width=0.85\columnwidth]{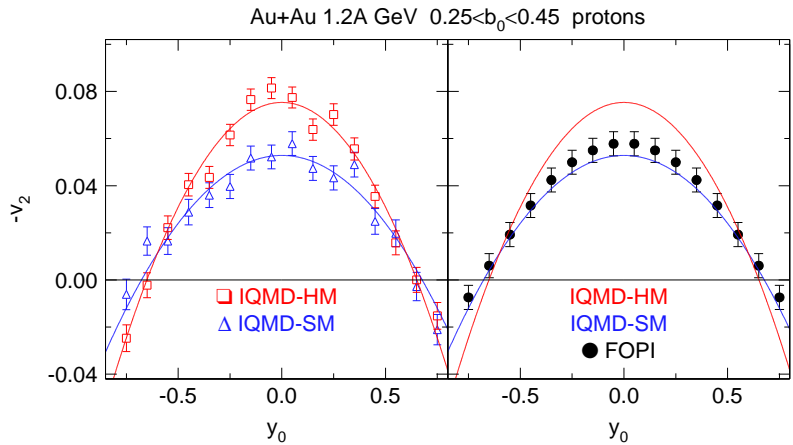}}
\vskip -0.1cm
\caption{Elliptic flow parameter -v2 (note the change of sign) of protons as a function of the normalized rapidity $y_0$
for $^{197}$Au + $^{197}$Au collisions at 1.2 GeV/nucleon and the indicated near-central interval of
normalized impact parameters $b_0$. Left panel: IQMD calculations (symbols) for a hard (HM) and a soft (SM)
EoS, both with momentum dependent forces, and the fit results (lines) assuming a quadratic dependence on $y_o$.
Right panel: The obtained fit results (lines) in comparison with the experimental data (symbols) measured
with the FOPI detector at the GSI laboratory
(reprinted from Ref.~8, Copyright (2016), with permission from Elsevier).
}
\label{fig:lefevre}
\end{figure}

The precision in interpreting the measured particle flows has been demonstrated by the FOPI Collaboration in their
extensive report by Reisdorf {\it et al.}\cite{reisdorf12} Calculations with the IQMD model were shown to account 
not only for the overall strength of the flow but also for the detailed dependence on rapidity and the change of sign
of the $v_2$ parameter at rapidities $|y_0| \approx 0.7$; here $y_0$ is the rapidity normalized to the projectile 
rapidity in the c.m. system (Fig.~\ref{fig:lefevre}). The approximately quadratic dependence of $v_2 (y_0)$ has 
been fitted with two parameters, $v_2 (y_0) = v_{20} + v_{22} \cdot y_0^2$, and a new quantity $v_{2n} = |v_{20}| +|v_{22}|$ 
has been introduced. It combines the information contained in the amplitude and the rapidity dependence of $v_2$. 
The dependence of $v_{2n}$ on the incident energy in the interval 0.4 to 1.5 GeV/nucleon covered by FOPI is fairly 
flat and its discriminating power between the soft and stiff parametrizations of the symmetric-matter EoS appears
rather convincing.\cite{reisdorf12,lefevre16}

\section{Results from FOPI-LAND}

The squeeze-out of neutrons has first been observed by the FOPI-LAND Collaboration who
studied the reaction $^{197}$Au + $^{197}$Au at 400 MeV/nucleon.\cite{leif93}
The squeeze-out of charged particles reaches its maximum at this energy (Fig.~\ref{fig:v2corr}),
and similarly large anisotropies were observed for neutrons.\cite{lamb94}
The neutrons had been detected with the Large-Area Neutron Detector, LAND,\cite{LAND} 
while the FOPI Forward
Wall, covering the forward hemisphere of laboratory angles $\theta_{\rm lab} \le 30^{\circ}$ 
with more than 700 plastic scintillator elements, was used to determine the
modulus and azimuthal orientation of the impact parameter.

The motivation for returning to the existing data set has been provided by studies 
performed with the UrQMD transport code for this fairly neutron-rich system ($N/Z = 1.49$). 
They indicated a significant sensitivity of the elliptic-flow parameters to the 
assumptions made for the density dependence of the symmetry energy.\cite{russotto11} 
In calculations with power-law coefficients $\gamma = 0.5$ and 1.5 
(cf. Eq.~\ref{eq:pot_term} and Fig.~\ref{fig:params}), 
the relative strengths of neutron
and proton elliptic flows were found to vary on the level of 15\%.

\begin{figure}[!htb]   %Fig. 7
\centerline{\includegraphics[width=0.66\columnwidth]{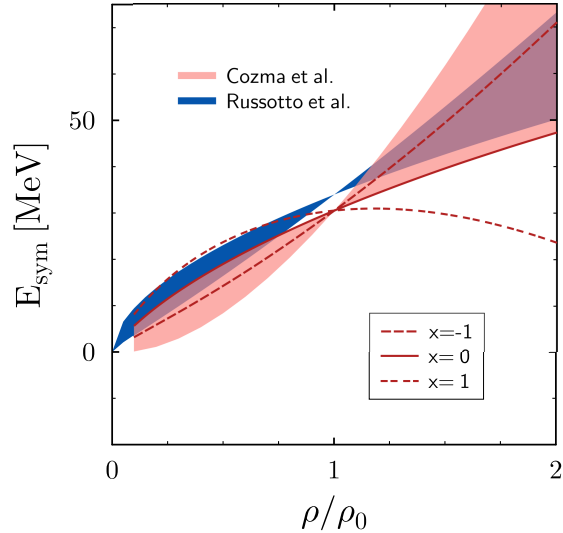}}
\vskip -0.1cm  
  \caption{Constraints on the density dependence of the symmetry energy obtained by Cozma {\it et al.} from 
comparing predictions of the T\"{u}bingen QMD for the neutron-proton elliptic-flow difference and ratio
to FOPI-LAND experimental data (Ref.~108). 
The result of Russotto {\it et al.} (Ref.~37) is also shown, 
together with the Gogny-inspired parametrization of the symmetry energy for three values of 
the stiffness parameter: x = -1 (stiff), x = 0, and x = 1 (soft)
(from Ref.~108).
}
\label{fig:cozma}
\end{figure}

The reanalysis of the data consisted mainly in choosing 
equal acceptances for neutrons and hydrogen isotopes with regard to 
particle energy, rapidity and transverse momentum (energy and momentum 
per nucleon for deuterons and tritons). 
The theoretical predictions have been obtained simulating the LAND acceptance and the 
experimental analysis conditions and were found to follow qualitatively the experimental data.

For the quantitative evaluation, the ratio of the flow parameters of neutrons versus protons 
or versus $Z=1$ particles has been proposed as a useful observable.\cite{russotto11}  
Systematic effects influencing the collective flows of neutrons and charged particles 
in similar ways should thereby be minimized, on the experimental as well as on the 
theoretical side. 
In consideration of the systematic and experimental errors, a value
$\gamma = 0.9 \pm 0.4$ has been adopted by the authors as best representing the power-law 
exponent of the potential term resulting from the elliptic-flow analysis.\cite{russotto11}  
It falls slightly below the linear $\gamma = 1.0$ line shown in Fig.~\ref{fig:params}.
The corresponding slope parameter is $L = 83 \pm 26$~MeV.
Comparing with the many-body
theories shown in Fig.~\ref{fig:fuchs06b}, the elliptic-flow result is in good qualitative 
agreement with the range spanned by the DBHF and variational calculations based on realistic
nuclear forces.

\begin{figure}[htb]   %Fig. 8
\centerline{\includegraphics[width=0.64\columnwidth]{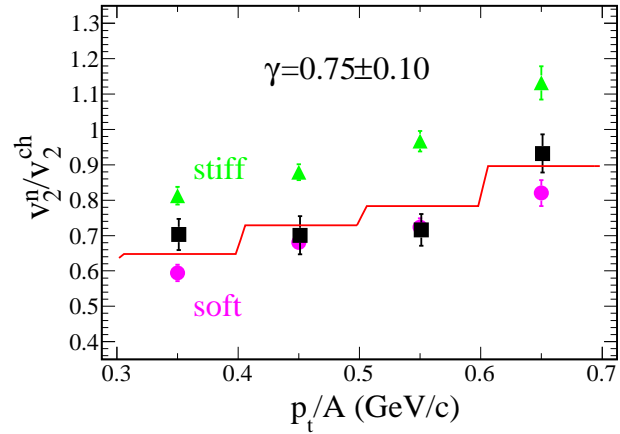}}
\vskip -0.1cm
\caption{Elliptic flow ratio of neutrons over all charged particles for central ($b<$ 7.5 fm) 
collisions of $^{197}$Au+$^{197}$Au at 400 MeV/nucleon
as a function of the transverse momentum per nucleon $p_{t}/A$.
The black squares represent the experimental data; the green triangles and purple circles represent the UrQMD predictions
for stiff ($\gamma =1.5$) and soft ($\gamma =0.5$) power-law exponents of the potential term, respectively. 
The solid line is the result of a linear interpolation between the predictions, 
weighted according to the experimental errors of the included four bins in $p_{t}/A$ and leading to the 
indicated $\gamma =0.75 \pm 0.10$
(from Ref.~51).
}
\label{fig:diffdata}
\end{figure}

In an independent analysis, Cozma has used data from the same experiment and investigated 
the influence of several parameters on the difference and ratio
of the elliptic flows of protons and neutrons 
using the T\"{u}bingen version of the QMD transport model.\cite{cozma13,russotto_epja14} They included the
parametrization of the isoscalar EoS, the choice of various forms of free or in-medium
nucleon-nucleon cross sections, and model parameters as, e.g., the widths of the wave 
packets representing nucleons. 
The interaction developed by Das {\it et al.} with an explicit 
momentum dependence of the symmetry energy part was used.\cite{das03,lichen05} 
In Fig.~\ref{fig:cozma}, the obtained constraint on the density dependence of the symmetry energy is presented. 
It is overall in agreement with the result reported by Russotto {\it et al.}\cite{russotto11} but has a larger 
uncertainty because the effects of exploring the full intervals of the above-mentioned  
theoretical parameters were taken into account in the error determination. 
A stiffness closely corresponding to the $x=-1$ scenario is clearly favored while a super-soft solution with $x=1$ is ruled out.

\section{Results from ASY-EOS}

The experimental setup of the ASY-EOS experiment at the GSI laboratory followed the scheme developed for FOPI-LAND by using the 
Large Area Neutron Detector (LAND\cite{LAND}) as the main instrument for neutron and charged particle detection. 
For the event characterization and for measuring the orientation of the reaction plane, three
detection systems had been installed. The ALADIN Time-of-Flight (AToF) Wall\cite{schuettauf96} was used 
to detect charged particles and fragments in forward direction at polar angles up to
$\theta_{\rm lab} \le 7^{\circ}$. Its capability of identifying large fragments and of characterizing
events with a measurement of $Z_{\rm bound}$\cite{schuettauf96} permitted the sorting of events 
according to impact parameter. Four double rings of the 
CHIMERA\cite{Pag04} multidetector
carrying together 352 CsI(Tl) scintillators in forward direction and four rings with 50 thin CsI(Tl) 
elements of the Washington University Microball\cite{muball} array surrounding the target
provided sufficient coverage and granularity for determining the orientation 
of the reaction plane from the measured azimuthal particle distributions. A detailed description of the experiment 
is available in Ref.~51.

\begin{figure}[htb!]   %Fig. 9
\centerline{\includegraphics[width=0.62\columnwidth]{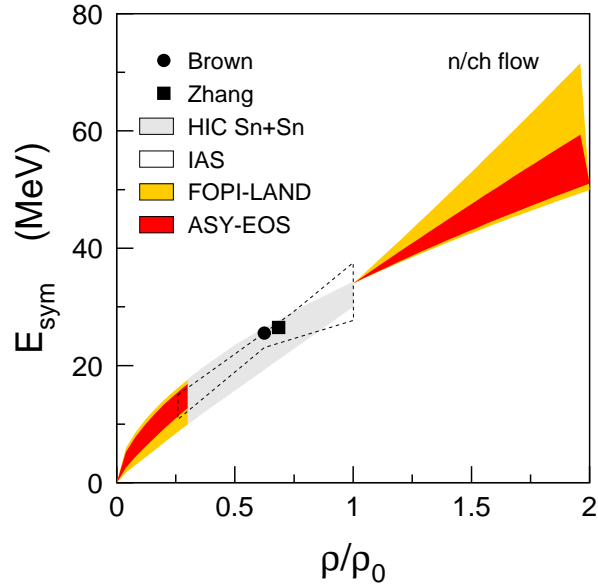}}
\vskip -0.1cm
\caption{Constraints deduced for the density dependence of the symmetry energy from the ASY-EOS data
in comparison with the FOPI-LAND result of Ref.~37 as a function of the reduced density $\rho/\rho_0$. 
The low-density results of Refs.~14, 34, 35 and 113 as reported in Ref.~16 are 
given by the symbols, the gray area (HIC), and the dashed contour (IAS). For clarity, the FOPI-LAND and ASY-EOS results are
not displayed in the interval $0.3 < \rho/\rho_0 < 1.0$
(from Ref.~51).
}
\label{fig:final}
\end{figure}

The ratio $v_2^{n}/v_2^{ch}$ obtained for the class of central ($b<$ 7.5 fm) 
collisions as a function of the transverse momentum per nucleon $p_{t}/A$ is shown in 
Fig.~\ref{fig:diffdata}. 
The best description of the neutron-vs-charged-particle elliptic flow is
obtained with a power-law coefficient $\gamma=0.75\pm 0.10$, where $\Delta\gamma=0.10$ is the statistical
uncertainty returned by the fit routine. It results from linearly interpolating between 
the predictions for the soft, $\gamma=0.5$, and the stiff, $\gamma=1.5$, predictions of the model
within the range of transverse momentum $0.3 \le p_{t}/A \le 0.7$~GeV/$c$.

With all corrections and errors included, the acceptance-integrated elliptic-flow ratio leads to a power-law coefficient $\gamma = 0.72 \pm 0.19$. 
This is the result displayed in Fig.~\ref{fig:final} as a function of the reduced density $\rho/\rho_0$. The new result confirms the former and has a 
considerably smaller uncertainty. It is also worth noting that the present parametrization is compatible with the low-density behavior of the symmetry 
energy from Refs.~14, 34, 35 and 113 that are included in the figure.  
The corresponding slope parameter describing the variation of the symmetry energy with density at saturation
is $L = 72 \pm 13$~MeV. 

The effective density probed with the elliptic flow measurement was explored in a study using again the 
T\"{u}bingen version\cite{cozma13} of the QMD model. The underlying idea
consisted in performing transport calculations for the present reaction with two parametrizations of the symmetry energy that were chosen
to be different for a selected range of density and identical elsewhere. The magnitude of the obtained difference between the two 
predictions for the elliptic flow ratio is considered a measure of the sensitivity to the selected density region.

It was found that the sensitivity achieved with the elliptic-flow ratio of neutrons over charged particles, 
the case chosen for the ASY-EOS analysis,  reaches its maximum close to saturation density but extends beyond 
twice that value. Interestingly, the sensitivity of the neutron-vs-proton flow ratio has its maximum in the 
1.4--1.5 $\rho_0$ region, i.e. at 
significantly higher densities than with light complex particles being included.
This observation contains an important potential for future experiments. With efficient isotope separation, 
flow measurements may give access to the curvature of the symmetry energy at saturation, in addition to the slope.

\section{Conclusion and Outlook}

The "flow of nuclear matter out of the regions of compressed densities" studied by Walter Greiner and his group\cite{greiner3} many years 
ago has, in the meanwhile, evolved into a useful tool for the study of the nuclear equation of state (EoS) at high density. 
According to the predictions of transport models, the relative strengths of
neutron and proton elliptic flows represent an observable sensitive to the symmetry
energy at densities near and above saturation. By forming ratios or differences
of neutron versus charged-particle flows, the influence of isoscalar-type
parameters of the model descriptions can be minimized.
As the interpretation of the FOPI pion ratios\cite{reis07} is not yet conclusive,  
this observable is presently unique as a terrestrial source of information for the EoS of asymmetric matter at high densities.

The value $\gamma = 0.72 \pm 0.19$ obtained in the ASY-EOS experiment for the power-law coefficient of the potential part 
in the UrQMD parametrization of the symmetry energy and the slope parameter $L = 72 \pm 13$~MeV are equivalent to a symmetry pressure 
$p_0 = \rho_0 L/3 = 3.8 \pm 0.7$~MeVfm$^{-3}$. The latter may be used to estimate the pressure in neutron-star matter at 
saturation density. For an assumed
asymmetry $\delta = (\rho_n - \rho_p)/\rho = 0.9$ in that part of the star, it amounts to 3.4~MeVfm$^{-3}$ (Ref.~51), 
a value that compares well with the pressure obtained by 
Steiner {\it et al.}\cite{steiner13} from neutron-star observations.
  
The study of the parameter dependence has shown that the contributions of the isoscalar sector are still significant. 
It will, therefore, be important to improve the description of the nuclear interaction in transport models,
to reduce the parameter ranges also in the isoscalar sector, to improve the algorithms used for clusterization,
as well as going beyond the mean-field picture, including short-range correlations.\cite{yong16} 
Moreover, it will be quite important to compare the experimental data with the predictions of several transport models, 
of both Boltzmann-Vlasov and molecular-dynamics type, in order to pursue the work towards a 
model-independent constraint of the high-density symmetry energy initiated in Refs.~108 and 109.

The presented experimental results and the theoretical study of the density range that has been probed, provide a strong 
encouragement for continuing the flow measurements with improved detection systems. Model 
studies indicate that the sensitivity to the stiffness of the symmetry energy is still significant at incident 
energies as high as 800 MeV or 1 GeV/nucleon. 
The density study suggests that the range of densities that can be probed in the laboratory may reach up to twice or 
three times the saturation value if higher precision and isotopic resolution for light charged particles can be achieved.  
Future experiments will, therefore, benefit from the improved capabilities of the NeuLAND detector\cite{NeuLAND} 
presently constructed as part of the $R^{3}B$ setup for experiments at FAIR. 

{\bf Acknowledgments}

Stimulating and fruitful discussions with M.~Di~Toro, W.~Reisdorf, Yongjia Wang, and with the authors of 
Ref.~109, M.~D.~Cozma, A.~Le~F\`{e}vre, Y.~Leifels, Qingfeng~Li, J.~{\L}ukasik, and P.~Russotto 
are gratefully acknowledged. HHW is partially supported by the Universe cluster of excellence of the DFG Germany.

%\begin{thebibliography}{0}
%\begin{thebibliography}{9}   % Use for  1-9  references


\begin{thebibliography}{99} % Use for 10-99 references
%\itemsep -2pt 

\bibitem{greiner1} M. Danos and W. Greiner, {\em Phys. Rev.} {\bf 134}, B284 (1964). 

\bibitem{greiner2} R. Ligensa, W. Greiner and M. Danos, {\em Phys. Rev. Lett.} {\bf 16}, 364 (1966). 

\bibitem{greiner4} 
W. Scheid, H. M\"{u}ller and W. Greiner, 
{\em Phys. Rev. Lett.} {\bf 32}, 741 (1974). 

\bibitem{dani02} 
P.~Danielewicz, R. Lacey and W. G. Lynch, 
{\em Science} {\bf 298}, 1592 (2002).
 
\bibitem{sturm01} 
C. Sturm {\it et al.}, %[KaoS Coll.],  
{\em Phys. Rev. Lett.} {\bf 86}, 39 (2001). 

\bibitem{fuchs01} 
C. Fuchs, A. Faessler, E. Zabrodin and Yu-Ming Zheng,   
{\em Phys. Rev. Lett.} {\bf 86}, 1974 (2001). 

\bibitem{reisdorf12} 
W. Reisdorf {\it et al.}, 
{\em Nucl. Phys. A} {\bf 876}, 1 (2012).

\bibitem{lefevre16} 
A.~Le F\`{e}vre, Y.~Leifels, W.~Reisdorf, J.~Aichelin and Ch.~Hartnack, 
{\em Nucl. Phys. A} {\bf 945}, 112 (2016).

\bibitem{greiner3} 
W. Scheid, R. Ligensa and W. Greiner, 
{\em Phys. Rev. Lett.} {\bf 21}, 1479 (1968). 

\bibitem{lattprak07}
J. M. Lattimer and M. Prakash, 
{\em Phys. Rep.} {\bf 442}, 109 (2007).

\bibitem{lipr08}
Bao-An Li, Lie-Wen Chen and Che Ming Ko, 
{\em Phys. Rep.} {\bf 464}, 113 (2008).

\bibitem{ditoro10}
M. Di Toro, V. Baran, M. Colonna and V. Greco,
{\em J. Phys. G} {\bf 37}, 083101 (2010).  

\bibitem{EPJA2014}
Topical Issue on Nuclear Symmetry Energy,
{\em Eur. Phys. Jour. A} {\bf 50}, number 2 (2014).

\bibitem{tsang09}
M.~B.~Tsang, Yingxun~Zhang, P.~Danielewicz, M.~Famiano, Z.~Li, W.~G.~Lynch and A.~W.~Steiner,
{\em Phys. Rev. Lett.} {\bf 102}, 122701 (2009).

\bibitem{tsang12}
M. B.~Tsang {\it et al.}, 
{\em Phys. Rev. C} {\bf 86}, 015803 (2012).

\bibitem{horo14}
C. J. Horowitz, E. F. Brown, Y. Kim, W. G. Lynch, R. Michaels, A. Ono, J. Piekarewicz, M. B. Tsang and H. H. Wolter,
{\em J. Phys. G} {\bf 41}, 093001 (2014).

\bibitem{horo06}
C. J. Horowitz and A. Schwenk, 
%{\it Phys. Lett. B} {\bf 638} (2006) 153.
{\em Nucl. Phys. A} {\bf 776}, 55 (2006).

\bibitem{nato10}
J. B. Natowitz {\it et al.},
{\em Phys. Rev. Lett.} {\bf 104}, 202501 (2010). 

\bibitem{typel10}
S. Typel, G. R\"{o}pke, T. Kl\"{a}hn, D. Blaschke and H. H. Wolter,
{\em Phys. Rev. C} {\bf 81}, 015803 (2010).

\bibitem{qin12}
L. Qin {\it et al.},
{\em Phys. Rev. Lett.} {\bf 108}, 172701 (2012). 

\bibitem{brown00}
B. A. Brown, 
{\em Phys. Rev. Lett.} {\bf 85}, 5296 (2000).

\bibitem{fuchs06}
C. Fuchs and H. H. Wolter, 
{\em Eur. Phys. J. A} {\bf 30}, 5 (2006).

\bibitem{subedi08}
R. Subedi {\it et al.},
{\em Science} {\bf 320}, 1476 (2008). 

\bibitem{hen14sc}
O.~Hen {\it et al.},
{\em Science} {\bf 346}, 614 (2014). 

\bibitem{xuli10}
Chang Xu and Bao-An Li,
{\em Phys. Rev. C} {\bf 81}, 064612 (2010).

\bibitem{steiner12} A. W. Steiner and S. Gandolfi, 
{\em Phys. Rev. Lett.} {\bf 108}, 081102 (2012). 
%arxiv:1110.4142v2

\bibitem{carb11}
A. Carbone, A. Polls and A. Rios,
{\em Europhys. Lett.} {\bf 97}, 22001 (2012).
%preprint arXiv:1111.0797[nucl-th] (2011).

\bibitem{hen15}
Or Hen, Bao-An Li, Wen-Jun Guo, L. B. Weinstein and E. Piasetzky,
{\em Phys. Rev. C} {\bf 91} 025803 (2015).

\bibitem{yong16}
Gao-Chan~Yong,
{\em Phys. Rev. C} {\bf 93}, 044610 (2016).

\bibitem{lihan13}
Bao-An Li and Xiao Han,
{\em Phys. Lett. B} {\bf 727}, 276 (2013). 

\bibitem{baran05}
V. Baran, M. Colonna, V. Greco and M. Di Toro,
{\em Phys. Rep.} {\bf 410}, 335 (2005).

\bibitem{klimk07}
A.~Klimkiewicz {\it et al.}, 
{\em Phys. Rev. C} {\bf 76}, 051603(R) (2007).

\bibitem{tamii11}
A. Tamii {\it et al.}, 
{\em Phys. Rev. Lett.} {\bf 107}, 062502 (2011).

\bibitem{zhangchen13}
Zhen Zhang and Lie-Wen Chen,
{\em Phys. Lett. B} {\bf 726}, 234 (2013). 

\bibitem{brown13}
B. A. Brown, 
{\em Phys. Rev. Lett.} {\bf 111}, 232502 (2013).

\bibitem{li02}
Bao-An Li, 
{\em Phys. Rev. Lett.} {\bf 88}, 192701 (2002).

\bibitem{russotto11}
P. Russotto {\it et al.}, 
{\em Phys. Lett. B} {\bf 697}, 471 (2011). 

\bibitem{leif93}
Y. Leifels {\it et al.}, 
{\em Phys. Rev. Lett.} {\bf 71}, 963 (1993).

\bibitem{lamb94}
D. Lambrecht {\it et al.}, 
{\em Z. Phys. A} {\bf 350}, 115 (1994).

\bibitem{rev_epja}
W. Trautmann and H. H. Wolter, 
{\em Int. J. Mod. Phys. E} {\bf 21}, 1230003 (2012).

\bibitem{reis07}
W. Reisdorf {\it et al.}, 
{\em Nucl. Phys. A} {\bf 781}, 459 (2007).

\bibitem{ferini05}
G. Ferini {\it et al.}, 
{\em Nucl. Phys. A} {\bf 762}, 147 (2005);
{\em Phys. Rev. Lett.} {\bf 97}, 202301 (2006).

\bibitem{xiao09}
Zhigang~Xiao {\it et al.}, 
{\em Phys. Rev. Lett.} {\bf 102}, 062502 (2009).

\bibitem{feng10}
Zhao-Qing Feng and Gen-Ming Jin, 
{\em Phys. Lett. B} {\bf 683}, 140 (2010).

\bibitem{xie13}
Wen-Jie Xie, Jun Su, Long Zhua and Feng-Shou Zhang
{\em Phys. Lett. B} {\bf 718}, 1510 (2013).

\bibitem{baoan11}
Bao-An Li {\it et al.},
{\em J. Phys. Conf. Ser.} {\bf 312}, 042006 (2011).

\bibitem{wen09}
De-Hua Wen, Bao-An Li and Lie-Wen Chen, 
{\em Phys. Rev. Lett.} {\bf 103}, 211102 (2009).

\bibitem{cozma11}
M. D. Cozma, 
{\em Phys. Lett. B} {\bf 700}, 139 (2011). 

\bibitem{LAND}
Th. Blaich {\it et al.},
{\em Nucl. Instrum. Methods Phys. Res. A} {\bf 314}, 136 (1992).

\bibitem{CHIMERA}
A. Pagano {\it et al.}, 
{\em Nucl. Phys. A} {\bf 734}, 504 (2004).

\bibitem{russotto16}
P. Russotto {\it et al.}, 
{\em Phys. Rev. C} {\bf 94}, 034608  (2016).

\bibitem{baldo04}
M. Baldo, C. Maieron, P. Schuck and X. Vi\~{n}as,
{\em Nucl. Phys. A} {\bf 736}, 241 (2004).

\bibitem{fukukawa15} 
K. Fukukawa, M. Baldo, G. F. Burgio, L. Lo Monaco and H.-J. Schulze, 
{\em Phys. Rev. C} {\bf 92}, 065802 (2015).

\bibitem{drischler16} 
C. Drischler, A. Carbone, K. Hebeler and A. Schwenk.
{\em Phys. Rev. C} {\bf 94}, 054307 (2016).

\bibitem{wiringa02}
R. B. Wiringa and S. C. Pieper, 
{\em Phys. Rev. Lett.} {\bf 89}, 182501 (2002).

\bibitem{burgio08}
F. Burgio,
{\em J. Phys. G} {\bf 35}, 014048 (2008).

\bibitem{hebe10}
K. Hebeler, J. M. Lattimer, C. J. Pethick and A. Schwenk,
{\em Phys. Rev. Lett.} {\bf 105}, 161102 (2010).
%EFT up to normal nuclear density; effect of three-body forces

\bibitem{steiner10}
A. W. Steiner, J. M. Lattimer and E. F. Brown,
{\em ApJ.} {\bf 722}, 33 (2010).
%review in arXiv:1005.0811v2

\bibitem{NICER}
see, e.g., the NASA website for the Neutron star Interior Composition ExploreR Mission (NICER), 
https://heasarc.gsfc.nasa.gov/docs/nicer/

\bibitem{giordano10}
V. Giordano, M. Colonna, M. Di Toro, V. Greco and J. Rizzo,
{\em Phys. Rev. C} {\bf 81}, 044611 (2010).

\bibitem{feng12}
Zhao-Qing Feng,
{\em Phys. Lett. B} {\bf 707}, 83 (2012). 
%preprint, arXiv:1111.3590[nucl-th] (2011).

\bibitem{liguoli15}
Xiao-Hua Li, Wen-Jun Guo, Bao-An Li, Lie-Wen Chen, F. J. Fattoyev and W. G. Newton,
{\em Phys. Lett. B} {\bf 743}, 408 (2015). 
%preprint, arXiv:1111.3590[nucl-th] (2011).

\bibitem{coupland16}
D. D. S. Coupland {\it et al.}, 
{\em Phys. Rev. C} {\bf 94}, 011601(R) (2016).

\bibitem{zhang14}
Yingxun Zhang, M. B. Tsang, Zhuxia Li, Hang Liu, 
{\em Phys. Lett. B} {\bf 732}, 186 (2014). 

\bibitem{qli06}
Qingfeng Li, Zhuxia Li, S. Soff, M. Bleicher and H. St\"{o}cker, 
{\em J. Phys. G} {\bf 32}, 151 (2006); {\it ibid.} {\bf 32}, 407 (2006). 

\bibitem{das03}
C. B. Das, S. Das Gupta, C. Gale and Bao-An Li, 
{\em Phys. Rev. C} {\bf 67}, 034611 (2003).

\bibitem{lichen05}
Bao-An Li and Lie-Wen Chen,
{\em Phys. Rev. C} {\bf 72}, 064611 (2005).

\bibitem{wang14}
Yongjia Wang, Chenchen Guo, Qingfeng Li, Hongfei Zhang, Y. Leifels and W. Trautmann,  
{\em Phys. Rev. C} {\bf 89}, 044603 (2014).

\bibitem{guo12}
ChenChen Guo, YongJia Wang, QingFeng Li, W. Trautmann, Ling Liu and LiJuan Wu, 
%{\it Science China Series G} {\bf 55} (2012) 252.
{\em Science China Physics, Mechanics \& Astronomy} {\bf 55}, 252 (2012).

\bibitem{lattlim13}
J.~M.~Lattimer and Y.~Lim,
{\em Astrophys. J.} {\bf 771}, 51 (2013).
%preprint arXiv:1203.4286[nucl-th] (2012).

\bibitem{todd_piek05}
B. G. Todd-Rutel and J. Piekarewicz,
{\em Phys. Rev. Lett.} {\bf 95}, 122501 (2005). 

\bibitem{rocamaza11}
X. Roca-Maza, M. Centelles, X. Vi\~{n}as and M. Warda,
{\em Phys. Rev. Lett.} {\bf 106}, 252501 (2011).

\bibitem{typel01}
S. Typel and B. A. Brown,
{\em Phys. Rev. C} {\bf 64}, 027302 (2001). 

\bibitem{furn02} 
R. J. Furnstahl,
{\em Nucl. Phys. A} {\bf 706}, 85 (2002).

\bibitem{horo12}
C. J. Horowitz {\it et al.},
{\it Phys. Rev. C} {\bf 85}, 032501(R) (2012).
%preprint, arXiv:1202.1468[nucl-ex] (2012).

\bibitem{PREX12}
S. Abrahamyan {\it et al.},
{\em Phys. Rev. Lett.} {\bf 108}, 112502 (2012).

\bibitem{li_npa02}
Bao-An Li, 
{\em Nucl. Phys. A} {\bf 708}, 365 (2002).

\bibitem{junxu13}
Jun Xu, Lie-Wen Chen, Che Ming Ko Bao-An Li and Yu~Gang~Ma,
{\em Phys. Rev. C} {\bf 87}, 067601 (2013).

\bibitem{greco03} 
V. Greco, V. Baran, M. Colonna, M. Di Toro, T. Gaitanos and H. H. Wolter, 
{\em Phys. Lett. B} {\bf 562}, 215 (2003). 

\bibitem{xlopez07}
X.~Lopez {\it et al.}, 
{\em Phys. Rev. C} {\bf 75}, 011901(R) (2007).

\bibitem{cozma16}
M. D. Cozma, 
{\em Phys. Lett. B} {\bf 753}, 166 (2016). 

\bibitem{ikeno16}
N. Ikeno, A. Ono, Y. Nara and A. Ohnishi,
{\em Phys. Rev. C} {\bf 93}, 044612 (2016).

\bibitem{song15}
Taesoo Song and Che Ming~Ko,
{\em Phys. Rev. C} {\bf 91}, 014901 (2015).
%preprint arXiv:1403.7363 [nucl-th] (2014).

\bibitem{baoan15}
Bao-An Li,
{\em Phys. Rev. C} {\bf 92}, 034603 (2015).

\bibitem{hong14}
Jun Hong and P. Danielewicz,
{\em Phys. Rev. C} {\bf 90}, 024605 (2014).

\bibitem{cozma17}
M. D. Cozma, 
{\em Phys. Rev. C} {\bf 95}, 014601 (2017). 

\bibitem{tsang17}
M. B. Tsang {\it et al.}, 
{\em Phys. Rev. C} {\bf 95}, 044614  (2017).

\bibitem{sprit}
R. Shane {\it et al.},
{\em Nucl. Instrum. Methods Phys. Res. A} {\bf 784}, 513 (2015).

%\bibitem{yong06}
%Gao-Chan Yong, Bao-An Li and Lie-Wen Chen,
%{\em Phys. Rev. C} {\bf 74}, 064617 (2006).

\bibitem{bertsch84}
G. F. Bertsch, H. Kruse and S. Das Gupta, 
{\em Phys. Rev. C} {\bf 29}, 673 (1984).

\bibitem{kruse85}
H. Kruse, B. V. Jacak, J. J. Molitoris, G. D. Westfall and H.~St\"{o}cker, 
{\em Phys. Rev. C} {\bf 31}, 1770 (1985).

\bibitem{aichelin86} 
J. Aichelin and H. St\"{o}cker, 
{\em Phys. Lett. B} {\bf 176}, 14 (1986). 

\bibitem{bertsch88} 
G. F. Bertsch and S. Das Gupta, 
{\em Phys. Rep.} {\bf 160}, 189 (1988). 

\bibitem{aichelin91} 
J. Aichelin,
{\em Phys. Rep.} {\bf 202}, 233 (1991). 

\bibitem{junxu16}
Jun Xu {\it et al.}, 
{\em Phys. Rev. C} {\bf 93}, 044609 (2016).

\bibitem{gutbrod90} 
H. H. Gutbrod {\it et al.}, 
{\em Phys. Rev. C} {\bf 42}, 640 (1990).

\bibitem{ollitrault97} 
J.-Y. Ollitrault, preprint arXiv:nucl-ex/9711003 (1997).
%correct definition with Sum2vn...

\bibitem{poskanzer98} 
A. M. Poskanzer and S. A. Voloshin, 
{\em Phys. Rev. C} {\bf 58}, 1671 (1998).
%correct definition with Sum2vn...

\bibitem{abelev07}
B. I. Abelev {\it et al.},    %Star
{\em Phys. Rev. Lett.} {\bf 99}, 112301 (2007). 

\bibitem{fries08}
R. Fries, V. Greco and P. Sorensen, 
{\em Annu. Rev. Nucl. Part. Sci.} {\bf 58}, 177 (2008). 

\bibitem{adare12}           %Phenix
A. Adare {\it et al.}, 
{\em Phys. Rev. C} {\bf 85}, 064914 (2012).

\bibitem{heinz13}
U.~W.~Heinz and R.~Snellings,
{\em Annu. Rev. Nucl. Part. Sci.} {\bf 63}, 123 (2013).
%Therefore, the large elliptic flow observed at RHIC energies provides compelling evidence for strongly interacting matter that behaves like an almost perfect liquid (118, 124).

\bibitem{dani00}
P. Danielewicz,
{\em Nucl. Phys. A} {\bf 673}, 375 (2000) 375.
%that the elliptic flow at midrapidity exhibits a particularly strong sensitivity to the meanfield momentum dependence in midperipheral to peripheral collisions. A relatively weak sensitivity was found in these collisions to the incompressibility of nuclear matter.

\bibitem{andro06}
A. Andronic, J.~{\L}ukasik, W. Reisdorf and W. Trautmann,
{\em Eur. Phys. J. A} {\bf 30}, 31 (2006).

\bibitem{lukasik05} 
J. {\L}ukasik {\it et al.}, 
{\em Phys. Lett. B} {\bf 608}, 223 (2005).

\bibitem{andronic05} 
A. Andronic {\it et al.}, 
{\em Phys. Lett. B} {\bf 612}, 173 (2005).

\bibitem{pinkenburg99}
C. Pinkenburg {\it et al.}, 
{\em Phys. Rev. Lett.} {\bf 83}, 1295 (1999). 

\bibitem{bmunzinger98} 
P. Braun-Munzinger and J. Stachel, 
{\em Nucl. Phys. A} {\bf 638}, 3c (1998).

\bibitem{cozma13}
M. D.~Cozma, Y.~Leifels, W.~Trautmann, Q.~Li and P.~Russotto, 
{\em Phys. Rev. C} {\bf 88}, 044912 (2013). 

\bibitem{russotto_epja14}
P.~Russotto, M. D.~Cozma, A.~Le~F\`{e}vre, Y.~Leifels, R.~Lemmon, J. {\L}ukasik, Q.~Li and W.~Trautmann, 
{\em Eur. Phys. J. A} {\bf 50}, 38 (2014).

\bibitem{schuettauf96} % Tof-Wall
A. Sch\"{u}ttauf {\it et al.}, 
{\em Nucl. Phys. A} {\bf 607}, 457 (1996).

\bibitem{Pag04}
A. Pagano {\it  et al.}, 
{\em Nucl. Phys. A} {\bf 734}, 504 (2004).

\bibitem{muball}
D. G.~Sarantites {\it  et al.}, 
%D. G.~Sarantites, P.-F.~Hua, M.~Devlin, L. G.~Sobotka, J.~Elson, J. T.~Hood, D. R.~LaFosse, J. E.~Sarantites and M. R.~Maier, 
{\em Nucl. Instr. and Meth. A} {\bf 381}, 418 (1996).

\bibitem{dani14} 
P.~Danielewicz and J.~Lee,
{\em Nucl. Phys. A} {\bf 922}, 1 (2014).
%preprint arXiv:1307.4130 (2013).

\bibitem{steiner13}
A.~W.~Steiner, J.~M.~Lattimer and E.~F.~Brown,
{\em Astrophys. J. Lett.} {\bf 765}, L5 (2013).

\bibitem{NeuLAND}
NeuLAND Technical Design Report, submitted to FAIR (2011).


\end{thebibliography}
\end{document}